# Towards Routine AI-Based PET/CT and SPECT/CT Lesion Segmentation and Tracking in PSMA Theranostics

Fereshteh Yousefirizi[1,2]✉, Jean-Mathieu Beauregard[3,4,5], Arman Rahmim[1,2,6]

[1]Department of Basic and Translational Research, BC Cancer Research Institute, Vancouver, BC, Canada.
[2]Department of Radiology, University of British Columbia, Vancouver, BC, Canada
[3]Department of Radiology and Nuclear Medicine; and Cancer Research Centre, Université Laval, Quebec City, QC, Canada
[4]Oncology Axis, CHU de Québec - Université Laval Research Centre, Quebec City, QC, Canada
[5]Division of Nuclear Medicine, Department of Medical Imaging, CHU de Québec – Université Laval, Quebec City, QC, Canada
[6] Departments of Physics and Biomedical Engineering, University of British Columbia, Vancouver, BC, Canada

**Corresponding Author:**

Fereshteh Yousefirizi, PhD
BC Cancer Research Institute
675 West 10th Ave
Vancouver, BC, V5Z 1L3
Canada
Email: frizi@bccrc.ca

**Abstract:**

Quantitative molecular imaging is central to treatment response assessment in oncology, yet clinical practice remains largely dominated by patient-level or limited target-lesion criteria that ignore inter-lesion heterogeneity. This limitation is particularly important in prostate cancer, where PSMA PET/CT can reveal extensive skeletal and nodal metastatic disease that often evolves heterogeneously under therapy. Accurate and scalable lesion segmentation and tracking across serial PSMA PET/CT and post-therapy SPECT/CT scans is therefore essential for implementing emerging PSMA-specific response frameworks, such as RECIP 1.0, and for enabling lesion-level dosimetry in $^{177}$Lu-PSMA radiopharmaceutical therapies (RPTs).

This article examines clinical motivations, technical foundations, and future pathways for automated lesion tracking in prostate cancer imaging. We focus on the unique requirements introduced by PSMA PET/CT compared with FDG PET/CT and highlight the critical role of quantitative SPECT/CT in linking imaging-derived disease characterization with delivered therapeutic dose. Recent advances in AI-based segmentation and automated lesion matching now make scalable longitudinal lesion correspondence feasible, providing comprehensive infrastructure for standardized response assessment and personalized PSMA-based theranostics

## 1. From Patient-Level Assessment to Lesion-Wise Understanding

Medical imaging has long been central to evaluating treatment response in oncology [1], yet current clinical practice still reduces complex, heterogeneous metastatic disease into simplified patient-level labels such as "partial response," "stable disease," or "progressive disease." This compression is at odds with the biological reality that metastatic lesions often behave differently from one another. Inter-lesion heterogeneity, where individual metastases show divergent patterns of shrinkage, stability, or progression under the same therapy, has been increasingly recognized as a key determinant of resistance and poor outcome [2–4].

Total tumor volume (TTV), also known as total metabolic (molecular) tumor volume (TMTV) for $^{18}$F-FDG (or other) PET scans, as well as total lesion activity (TLA) and total lesion fraction (TLF), provide quantitative and biologically meaningful estimates of systemic disease burden that go beyond conventional voxel- or lesion-based metrics such as SUVmax [5,7,9] (Table 1).

In prostate cancer, PSMA PET–derived tumor volume has emerged as a robust imaging biomarker, correlating strongly with disease extent, progression-free survival, and overall survival across treatment settings, including hormone-sensitive and metastatic castration-resistant disease [6,8,10]. PSMA-based volumetric biomarkers (e.g. TTV, TLA and TLF) capture total tumor burden and target expression, and can outperform conventional staging and purely visual assessment in risk stratification and treatment selection [11,12]. Similar evidence exists for FDG-based TMTV in lymphoma and other malignancies, where volumetric PET biomarkers provide superior prognostic stratification compared with visually based response criteria such as Lugano or Deauville scores [13–18]. Despite this extensive clinical validation across radiopharmaceuticals and cancer types, routine clinical adoption of PET-derived TTV remains limited, largely due to the labor-intensive and variable nature of manual segmentation, reinforcing the need for reliable AI-driven automation to enable scalable, reproducible volumetric quantification in clinical workflows.

**Table 1:** From Metabolic to Volumetric Response Evaluation, PR: Partial Response PD: Progressive Disease, CR: complete response, SD: Stable Disease

| Criteria | Imaging Modality | What It Measures | Core Principle | Best For | Limitations |
|---|---|---|---|---|---|
| **RECIST 1.1** Response Evaluation Criteria in Solid Tumors) | CT / MRI | **Lesion size** (longest diameter) | Change in linear dimensions + new lesion | Anatomical tumor shrinkage (soft-tissue disease) | Fails in bone-dominant or molecularly heterogeneous disease; ignores non-measurable lesions |
| **PERCIST** (PET Response Criteria in Solid Tumors) | FDG PET/CT | **Metabolic activity (SULpeak)** | 30% threshold for SUV change+ new lesion | Early metabolic response (FDG-avid disease) | Designed for FDG; limited transfer to PSMA PET; focuses on single lesion or up to 5 targets, not total tumor volume |

| RECIP 1.0 (Response Evaluation Criteria in PSMA PET/CT) | PSMA PET/CT | Total Tumor Volume (TTV) + New Lesions | ≥20 % ↑ TTV + new lesion = PD ; ≥30 % ↓ TTV = PR | Long-term response / overall survival prediction | Requires volumetric segmentation; not yet standardized across centers |
|---|---|---|---|---|---|

Beyond PET imaging, SPECT-derived TTV has emerged as a clinically meaningful biomarker, particularly in the context of radiopharmaceutical therapy (RPT) response assessment. Quantitative post-therapy SPECT/CT enables estimation of TTV and lesion activity that reflect delivered radiation burden rather than static radiopharmaceutical uptake alone. In patients undergoing [$^{177}$Lu]Lu-PSMA therapy, changes in SPECT-derived TTV have repeatedly demonstrated prognostic relevance, with increasing or non-decreasing tumor volume associated with significantly shorter progression-free survival and poorer outcomes, even in cases where PSA reduction is observed [19]. Importantly, volumetric SPECT biomarkers consistently outperform uptake metrics such as SUVmax, which have shown limited or inconsistent association with survival endpoints [20]. Advances in quantitative SPECT/CT, including same-day post-infusion imaging and next-generation digital systems, have further strengthened the feasibility of TTV assessment, demonstrating that reductions in SPECT-derived tumor volume correlate strongly with both overall survival and PSA-based response, while changes in SUV metrics do not [21]. Collectively, these findings establish SPECT-based TTV as a robust, treatment-relevant biomarker that complements PET-derived volumetric measures, particularly for longitudinal response assessment and decision-making during RPTs.

Traditional response criteria, designed around measurements of a small number of "target lesions" [1], fail to capture the full complexity of disease evolution and inevitably overlook clinically meaningful differences in lesion-level behaviors that may result in mixed response (Panel 1, Figure 1). Lesion tracking offers a solution by establishing correspondence between individual metastases across serial imaging time points. Through this longitudinal alignment, lesion tracking makes it possible to quantify how each metastasis evolves over time, capturing volumetric change, variations in radiopharmaceutical uptake, and the emergence of new lesions (Panel 2, Figure 1). Despite these advances, SPECT imaging, particularly post-therapy SPECT/CT in RPTs, remains comparatively underexplored [22], even though it is essential for assessing delivered dose and real-time treatment effect.

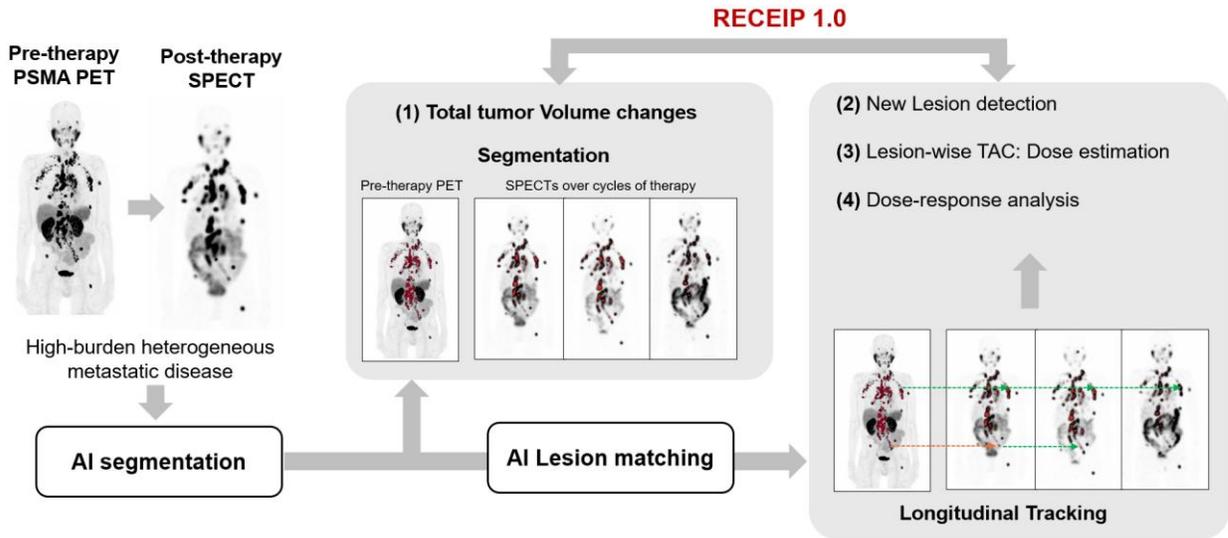

**Figure 1.** Proposed extension in the favor of RECEIP 1.0: incorporation of lesion tracking across time points and therapy cycles, enabling consistent lesion correspondence and detection of new lesions. This also supports reliable lesion-level TAC fitting, TIA assessment, and absorbed dose calculation, thereby advancing individualized dose–response evaluation in radiopharmaceutical therapies.

Prostate cancer, and especially PSMA-based imaging and therapy, highlights the urgency of moving beyond soft-tissue-restricted anatomical size measurements. This is particularly critical in metastatic castration-resistant prostate cancer (mCRPC), where bone-dominant disease is common and conventional structural criteria such as RECIST 1.1 are inadequate for characterizing sclerotic lesions. PSMA PET transforms previously "non-measurable" metastases into quantifiable targets, enabling a far more detailed understanding of disease biology. In addition, responding bone lesions may not always be associated with "shrinkage" (i.e. decrease in the extent of uptake) on PET or SPECT, and volumetric metrics other than TTV, such as TLA or TLF, may perform better for response assessment of bone lesions.

The emergence of PSMA-specific quantitative frameworks reflects this shift. Efforts such as PROMISE/miTNM [23] and the Standardised PSMA PET/CT Analysis and Reporting Consensus (SPARC) [24] consensus focus on standardized, reproducible quantification of PSMA expression. TTV derived from PSMA PET/CT has emerged as a powerful prognostic biomarker. RECIP 1.0, the systematic PSMA PET/CT response framework, recognizes whole-body TTV and new-lesion appearance as the core elements of clinically meaningful progression. These frameworks fundamentally rely on the ability to compare lesions across time, making robust lesion tracking essential for their consistent implementation. Without lesion-level longitudinal correspondence, the intended quantitative precision of PSMA imaging remains unattainable (Table 2).

**Table 2:** RECIP 1.0 Categories, PR: Partial Response PD: Progressive Disease, CR: complete response, SD: Stable Disease

| Category | Criteria |
|---|---|
| **RECIP-CR** | Absence of any PSMA uptake on follow-up PET scan. |
| **RECIP-PR** | >30% decrease in PSMA-derived tumor volume (PSMA-VOL) without appearance of new lesions. |
| **RECIP-PD** | ≥20% increase in PSMA-VOL with appearance of new lesions. |
| **RECIP-SD** | <30% decrease in PSMA-VOL with or without appearance of new lesions, or ≥30% decrease in PSMA-VOL with appearance of new lesions, or <20% increase in PSMA-VOL with/without new lesions, or ≥20% increase in PSMA-VOL without new lesions. |

## *2. Why PET/CT Requires Lesion Tracking*

FDG PET/CT has historically guided the development of PET-based response criteria, but PSMA imaging introduces fundamentally different clinical priorities. FDG-based frameworks such as PERCIST rely heavily on SUV changes in a small number of dominant lesions (24). In prostate cancer, FDG PET is increasingly recognized as complementary, particularly in revealing biologically aggressive disease that may be occult on PSMA imaging. PSMA PET/CT itself captures a broad landscape of disease, especially extensive skeletal metastases, and provides biologically relevant information about receptor expression beyond metabolic activity. Dual FDG–PSMA imaging has highlighted pronounced intra-patient heterogeneity, most notably in bone lesions, where discordant tracer uptake reflects divergent tumor phenotypes [25]. This receptor- and metabolism-based dissociation introduces new dimensions of heterogeneity, with lesions within the same patient exhibiting variable PSMA expression and FDG avidity independent of size, revealing the limitations of single-tracer, target-lesion–based response frameworks.

FDG-guided response paradigms primarily based on uptake intensity variations therefore cannot simply be transplanted into PSMA imaging. In prostate cancer, disease is often spatially widespread and highly heterogeneous, making lesion-level analysis indispensable for identifying patterns of resistance, dedifferentiation, and clonal escape. Lesion-level quantification also matters directly for therapy selection. Decisions regarding $^{177}$Lu-PSMA therapy increasingly rely not only on the presence of PSMA-avid disease but on its extent, intensity, and distribution across lesions, as well as the absence of FDG-positive PSMA-negative lesions that confer an adverse prognosis while not being targeted by the RPT. Whole-body TTV and PSMA expression metrics enable reproducible, quantitative assessments that outperform isolated SUV measurements in prognostic performance.

In this setting, lesion tracking becomes a structural necessity. Volume-based criteria such as those in RECIP 1.0 [26] rely on consistency in measuring TTV over time. New-lesion detection is a key determinant of progression, and longitudinal correspondence between lesions ensures that newly appearing metastases are correctly identified rather than confused with morphological changes of pre-existing ones. Tracking also enables more sophisticated metrics of response, including quantifying which lesions shrink most, which remain stable, and which progress despite therapy, all crucial for understanding therapeutic resistance.

*3. Integrating PET/CT and SPECT/CT: Lesion Tracking as the Backbone of Theranostics*

PSMA PET/CT provides high-resolution, receptor-targeted imaging for staging, treatment selection, and baseline quantification. SPECT/CT, especially quantitative SPECT acquired after $^{177}$Lu-PSMA therapy, complements this by visualizing actual radiopharmaceutical uptake and clearance patterns in vivo. The two modalities represent the diagnostic and therapeutic "bookends" of the theranostic paradigm. Lesion tracking is the computational infrastructure that connects them.

Post-therapy SPECT/CT provides the data required for dosimetry, estimating time-activity curves (TACs), time-integrated activity (TIA), and absorbed dose, for individual lesions [27]. These calculations require consistent lesion identification across multiple SPECT time points within each therapy cycle and across cycles. Without automated lesion tracking, current dosimetry workflows may restrict analysis to a few manually selected lesions, ignoring the majority of disease burden. This dramatically limits the ability to conduct lesion-level dose–response studies, which are essential for optimizing RPT regimens, understanding resistance, and potentially adapting therapy mid-course.

Recent SPECT studies have shown that volumetric changes in SPECT-derived TTV after the second or third therapy cycle correlate strongly with PSA progression-free survival, and that increases in SPECT TTV (>20%) or the appearance of new SPECT-positive lesions predict substantially poorer outcomes [20]. Conversely, significant reductions in SPECT TTV (>30%) are associated with improved overall survival [21]. It is also shown that the appearance of new lesions on posttherapy SPECT/CT images after cycle 2 is an independent prognostic factor for overall survival [19]. These findings parallel the logic of RECIP 1.0 and further reinforce the clinical importance of volumetric and lesion-level assessment. Without automated longitudinal lesion segmentation and tracking, these metrics cannot be used at scale.

Thus, lesion tracking emerges as an indispensable bridge between PSMA PET/CT and SPECT/CT, enabling consistent TTV calculation, reliable new-lesion detection, and accurate lesion-wise dosimetry. It transforms qualitative or semi-quantitative interpretation into a reproducible, quantitative workflow that aligns with current and emerging clinical protocols.

*4. Technical Foundations: AI Segmentation and Lesion Tracking Methodologies*

Robust segmentation is the foundation of any quantitative disease burden or lesion tracking pipeline, as accurate estimation of TTV critically depends on reliable, reproducible delineation of all lesion extents. Manual segmentation is impractical in patients with high disease burden and becomes prohibitive when assessments must be performed longitudinally across multiple time points, contributing to substantial inter-observer variability and limiting clinical scalability [28,29]. Recent advances in deep learning have enabled fully automated segmentation frameworks for PET/CT imaging that directly support whole-body total tumor volume quantification across multiple malignancies, including prostate cancer imaged with PSMA PET/CT[30,31], lymphoma [35,37], melanoma [16] and lung [32,33] with FDG PET/CT, and neuroendocrine tumors with somatostatin receptor PET [31].

Contemporary segmentation networks increasingly incorporate multi-scale encoders, cascaded 2D/3D architectures, transformer-enhanced backbones, and radiomics-aware representations, achieving segmentation performance approaching inter-observer agreement while maintaining robustness across multi-center data and varying scanner protocols [34]. Beyond purely image-driven approaches, emerging report-informed and vision–language segmentation methods integrate radiology report semantics or clinical textual context into 3D segmentation models, anchoring segmentation outcomes to physician-described disease patterns and improving clinical coherence and interpretability of TTV estimates [35]. Collectively, these AI-driven segmentation advances enable automated, reproducible, and scalable TTV extraction, providing the technical foundation necessary for longitudinal lesion tracking, response-adaptive imaging biomarkers, and future physician-in-the-loop decision support frameworks.

On this foundation, automated lesion tracking can be performed using several established approaches. Registration-based methods rely on aligning sequential images and matching lesions via spatial overlap; while intuitive, these methods struggle with deformable anatomy, low-resolution SPECT data, and changes in lesion morphology [36]. Pairwise feature-based matching constructs a cost matrix from lesion characteristics, centroid distance, volume, uptake changes, and radiomics features, and solves an optimal assignment problem between lesions at consecutive time points [38,39]. These technical advancements make it feasible to build unified PET-PET, PET–SPECT and SPECT-SPECT tracking pipelines that handle segmentation, correspondence, and response quantification in a clinically aligned manner.

## 5. Toward Routine Clinical Integration

The case for routine lesion tracking in PSMA-targeted theranostics is compelling. PSMA PET/CT and post-therapy SPECT/CT imaging provides a quantitative imaging paradigm that fundamentally depends on consistent lesion identification over time. While FDG-based imaging in hematologic oncology and could often rely on uptake metrics of dominant lesion(s), prostate cancer, particularly in its advanced stages, cannot. Bone-dominant, heterogeneous, multi-focal disease requires lesion-level precision to support treatment decisions, prognostic assessments, and dose–response modeling. Volumetric criteria such as RECIP 1.0 and SPECT-derived TTV metrics point toward a future where lesion-level analysis is integral to clinical workflows.

AI-based segmentation and topology-aware lesion tracking enable lesion-level analysis by accounting for spatial location, heterogeneity, and distance maps and similarity conditions among lesions across the body, providing the tools needed to realize this vision. Their clinical adoption would make personalized dosimetry feasible at scale, improve response assessment, and enable more rational therapeutic adaptation. Importantly, lesion tracking should not be viewed as a supplementary research capability but as essential infrastructure for modern PSMA theranostics. It provides continuity across PET/CT and SPECT/CT scans, RPT cycles, and evolving disease patterns, allowing clinicians to understand not just whether a patient is responding, but how each metastasis behaves, which lesions drive progression, and where treatment resistance emerges. Looking ahead, the same lesion-level tracking paradigm could play a critical role in linking PSMA-based imaging with complementary FDG PET, enabling systematic characterization of phenotypic heterogeneity across tracers and over time, as increasingly demonstrated in dual-tracer prostate cancer imaging studies

As theranostics and RPTs expand globally, transitioning these methods into routine practice will be critical. Harmonized imaging protocols, integrated clinical interfaces, and prospective trials incorporating lesion-level metrics will accelerate adoption. Ultimately, implementing robust AI-based lesion tracking will transform prostate cancer imaging into a longitudinal, lesion-centric discipline that more accurately reflects the biological complexity of disease and improves the precision of therapeutic decision-making.

## *Conflicts of interest*

Arman Rahmim is co-founder of Ascinta Technologies. No other potential conflicts of interest are reported.

## *Acknowledgments*

This research was supported by the Canadian Institutes of Health Research (CIHR) Project Grants PJT-180251 and PJT-173231.